\documentclass[conference]{IEEEtran}
\IEEEoverridecommandlockouts
\usepackage{cite}
\usepackage{amsmath,amssymb,amsfonts}
\usepackage{algorithmic}
\usepackage{graphicx}
\usepackage{textcomp}
\usepackage{subfigure}
\usepackage{algorithm}
\usepackage{xcolor}
\usepackage{colortbl,booktabs}
\usepackage{cite}
\usepackage{amsmath}
\usepackage{threeparttable}  
\usepackage{booktabs}  
\usepackage{diagbox}
\usepackage{multirow}
\usepackage{multicol}
\usepackage{stfloats}
\usepackage{setspace}
\usepackage{fancyhdr}
\usepackage{bm}
\def\BibTeX{{\rm B\kern-.05em{\sc i\kern-.025em b}\kern-.08em
    T\kern-.1667em\lower.7ex\hbox{E}\kern-.125emX}}
\begin{document}

\title{A Robust Semantic Communication System for Image\\

}

\author{
	\IEEEauthorblockN{
	Xiang Peng\IEEEauthorrefmark{1},
	Zhijin Qin\IEEEauthorrefmark{1}\IEEEauthorrefmark{2},
	Xiaoming Tao\IEEEauthorrefmark{1},
	Jianhua Lu\IEEEauthorrefmark{1}, 
	Khaled B. Letaief\IEEEauthorrefmark{3}}

	\IEEEauthorblockA{\IEEEauthorrefmark{1}Department of Electronic Engineering, Tsinghua University, Beijing, China}
	\IEEEauthorblockA{\IEEEauthorrefmark{2}Beijing National Research Center for Information Science and Technology (BNRist), Beijing, China}
	\IEEEauthorblockA{\IEEEauthorrefmark{3}Department of Electronic and Computer Engineering, Hong Kong University of Science and Technology, Hong Kong, China}
	\IEEEauthorblockA{Email: px21@mails.tsinghua.edu.cn, qinzhijin@tsinghua.edu.cn, \{taoxm, lhh-dee\}@mail.tsinghua.edu.cn, eekhaled@ust.hk.}
	}

\maketitle

\begin{abstract}
Semantic communications have gained significant attention as a promising approach to address the transmission bottleneck, especially with the continuous development of 6G techniques. Distinct from the well investigated physical channel impairments, this paper focuses on semantic impairments in image, particularly those arising from adversarial perturbations. Specifically, we propose a novel metric for quantifying the intensity of semantic impairment and develop a semantic impairment dataset. Furthermore, we introduce a deep learning enabled semantic communication system, termed as DeepSC-RI, to enhance the robustness of image transmission, which incorporates a multi-scale semantic extractor with a dual-branch architecture for extracting semantics with varying granularity, thereby improving  the robustness of the system. The fine-grained branch incorporates a semantic importance evaluation module to identify and prioritize crucial semantics, while the coarse-grained branch adopts a hierarchical approach for capturing the robust semantics. These two streams of semantics are seamlessly integrated via an advanced cross-attention-based semantic fusion module. Experimental results demonstrate the superior performance of DeepSC-RI under various levels of semantic impairment intensity.
\end{abstract}

\begin{IEEEkeywords}
	Semantic communications, semantic impairments, image transmission, multi-scale Vision Transformer.
\end{IEEEkeywords}

\section{Introduction}

\IEEEPARstart{S}{emantic} communications have been treated as a promising technology to improve the transmission efficiency~\cite{qin_survey}. In contrast to traditional communications, semantic communications diverge from the conventional emphasis on character-level accuracy and instead prioritize the faithful conveyance of semantic information. By embracing this innovative optimization objective, semantic communications can effectively reduce the volume of data to be transmitted, thereby successfully mitigating the challenges arising from the rapid data growth within communication networks~\cite{zhg_survey}.

Semantics, as the transmitted content for semantic communications, are typically represented as the information that are relevent to the specific task at the receiver. This is benefited from the development of the deep neural network (DNN), which is indispensable for semantic communications. The story of semantic communication unfolds through a multitude of tasks that systems can accomplish.

Semantic communications are capable of transmitting single-modal data over hostile channels, such as text, speech, image, and video \cite{DeepSC,zhg,Speech,TCOM_semantic_image,fu2023vector,qin2023computing,semantic_video,cross-modal}. Xie \textit{et al}.~\cite{DeepSC} made significant contributions with their pioneering work in semantic communications, which performs robust text transmission over hostile channels. Lu \textit{et al}.~\cite{zhg} designed a confidence-based distillation mechanism for efficient semantic encoding and proposed a semantic text communication system by utilizing reinforcement learning to address the semantic gap. Weng \textit{et al}.~\cite{Speech} presented a speech semantic communication system, which reduces the data volume by representing speech semantics as text. Kang \textit{et al}.~\cite{TCOM_semantic_image} proposed a semantic communication framework relying on deep reinforcement learning to improve classification accuracy for downstream tasks. Fu \textit{et al}.~\cite{fu2023vector} devised a knowledge base aided semantic communication system for conducting image transmission. Qin \textit{et al}.~\cite{qin2023computing} exploited the computing networks enabled semantic communication system to overcome the transmission limitation. Jiang \textit{et al}.~\cite{semantic_video} proposed a semantic video conference system to reduce the transmission load by representing semantics of human face with keypoints. Xie \textit{et al}.~\cite{cross-modal} provided a detailed analysis of multi-modal data transmission and multi-task execution within the context of semantic communications.

Although semantic communications are capable of handling various modal data and completing corresponding downstream tasks, they could be susceptible to \textit{semantic impairments}~\cite{peng2022robust}, which are defined as the signals that introduce semantic mismatch between the transmitter and the receiver. For instance, an image with adversarial perturbations could confuse the classification model at the receiver and lead to semantic ambiguity to the system~\cite{adversarial-image}.

Semantic impairments significantly impact the fidelity of semantic communications, primarily due to the fact that semantic communications heavily rely on DNNs, which are inherently sensitive to disturbances~\cite{DNN-vulnerability}. To ensure reliable semantic communications, semantic impairments must be considered.

Distinct from the extensively investigated physical channel noise and fading effects, we elaborate on the mechanism of semantic impairments and establish a robust semantic communication system to fight against semantic impairments. Moreover, the proposed method eliminates the need of retraining downstream task models to against semantic impairments in contrast to the adversarial training methods~\cite{FGM}. This achievement is made possible by utilizing computational resources allocated for semantic communications to mitigate semantic impairments. The distinctive contributions of our work are further detailed in a point-wise manner.

\begin{itemize}
	\item We propose a novel metric termed as \textit{image semantic impairment intensity} for quantifying the intensity of semantic impairments.
	\item We construct an image semantic impairment dataset with varying levels of semantic impairments for assessing the robustness of systems.
	\item Moreover, we propose a deep learning enabled semantic communication system for robust image transmission, namely DeepSC-RI, which leverages the mutli-scale semantic information to substantially mitigate semantic impairments and enhance semantic fidelity.
\end{itemize}

The rest of this paper is organized as follows. Section \uppercase\expandafter{\romannumeral2} introduces the semantic communication system models with particular emphasis on semantic impairments. Section \uppercase\expandafter{\romannumeral3} presents our proposed robust semantic communication system design, while our experimental results are discussed in Section \uppercase\expandafter{\romannumeral4}. Finally, section \uppercase\expandafter{\romannumeral5} concludes this paper.

\section{Semantic Communication System Model}

\begin{figure*}[tbp]
	\centering
	\includegraphics[scale=0.3]{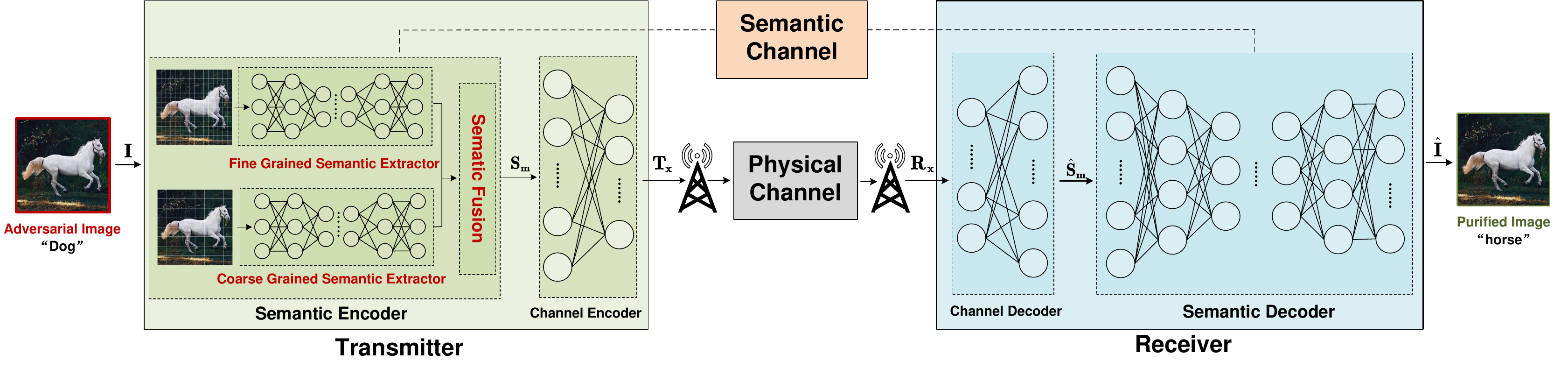}
	\caption{Overview of the proposed DeepSC-RI System.}
	\label{fig: system model}
\end{figure*}

As illustrated in Fig.~\ref{fig: system model}, we focus on an image transmission system with physical channels and semantic impairments. 

\subsection{Transmitter}

The corrupted image to be transmitted is denoted as $\bm{{\rm I}}$. The multi-scale semantic information, $\bm{{\rm S_m}}$, can be extracted with the robust semantic encoder, which can be represented as
\begin{equation}
\bm{{\rm S_m}} = f_s(\bm{{\rm I}}; \bm{\zeta}),
\end{equation}
where $f_s(\cdot; \bm{\zeta})$ is the robust semantic encoder having the trainable parameter set $\bm{\zeta}$.

Afterwards, the transmitted signal $\bm{{\rm T_x}}$ is obtained by conducting channel encoding, which is formulated as
\begin{equation}
\bm{{\rm T_x}} = f_c(\bm {{\rm S_m}}; \bm{\epsilon}),
\end{equation}
where $f_c(\cdot; \bm{\epsilon})$ is the channel encoder having the trainable parameter set $\bm{\epsilon}$. 

\subsection{Channel}

The received signal $\bm{{\rm R_x}}$ can be obtained as
\begin{equation}
\bm{{\rm R_x}} = \bm{{\rm H}} \cdot \bm{{\rm T_x}} + \bm{{\rm N}_p},
\end{equation}
where $\bm{{\rm H}}$ represents the coefficients of physical channels and $\bm{{\rm N}_p} \sim{\mathcal{CN} (0, \sigma_n^ 2)}$. In this paper, we consider the physical impairments imposed by AWGN and Rician fading channels.

\subsection{Receiver}

The received multi-scale semantic information, $\bm{{\rm \hat{S}_m}}$, is recovered after passing through the channel decoder, which can be represented as
\begin{equation}
\bm{{\rm \hat{S}_m}} = f^{-1}_c(\bm {{\rm R_x}}; \bm{\gamma }),
\end{equation}
where $f^{-1}_c(\cdot; \bm{\gamma})$ is the channel decoder having the trainable parameter set $\bm{\gamma}$.

The received image, $\bm{{\rm \hat{I}}}$, can be obtained by conducting semantic decoding, which is given by
\begin{equation}
\bm{{\rm \hat{I}}} = f^{-1}_s(\bm {{\rm \hat{S}_m }}; \bm{\varphi}),
\end{equation}
where $f^{-1}_s(\cdot; \bm{\varphi})$ is the semantic decoder having the trainable parameter set $\bm{\varphi}$.

The proposed semantic communication system is designed to fight against semantic impairments in image transmission, which is achieved by designing a robust semantic encoder and train the whole system jointly in an end-to-end manner. 

\subsection{\textit{Image Semantic Impairment Intensity}}

To quantitatively describe the intensity of semantic impairments in image, we propose \textit{image semantic impairment intensity} (ISII), which could be represented as
\begin{equation}
        \bm{{\rm ISII}} = 1 - \frac{{\bm V_\Phi}({\bm{{\rm I_u}}}) \cdot {\bm V_\Phi}({\bm{{\rm I_c}}})^T} {\Vert {\bm V_\Phi}({\bm{{\rm I_u}}}) \Vert \Vert {\bm V_\Phi}({\bm{{\rm I_c}}}) \Vert},
\end{equation}
where ${\bm V_\Phi}(\cdot)$ represents the function of the popular VGG Net~\cite{vgg}, which is a pretrained model with over 100 million parameters. ${\bm{{\rm I_c}}}$ is the corrupted image with semantic impairments, while ${\bm{{\rm I_u}}}$ is the corresponding uncorrupted image. 

\section{Proposed Robust Semantic Communication System for image}
In this section, we propose a deep learning enabled semantic communication system for robust image transmission with multi-scale semantic information.


\subsection{Multi-Scale Semantic Encoder}

While Vision Transformer (ViT)~\cite{vit} demonstrates superior performance over traditional Convolutional Neural Networks (CNNs) in many tasks, it processes images by dividing them into patches with a fixed-size, potentially hindering the capability of ViT to handle specific details and multi-grained features. 
To tackle this problem, we develop a multi-scale ViT as the backbone of the semantic encoder by incorporating modified ViT components in two branches, coupled with an efficient semantic fusion module.

\subsubsection{Fine-Grained Semantic Extractor}

\begin{figure}[tbp]
        \centering
        \includegraphics[scale=0.5]{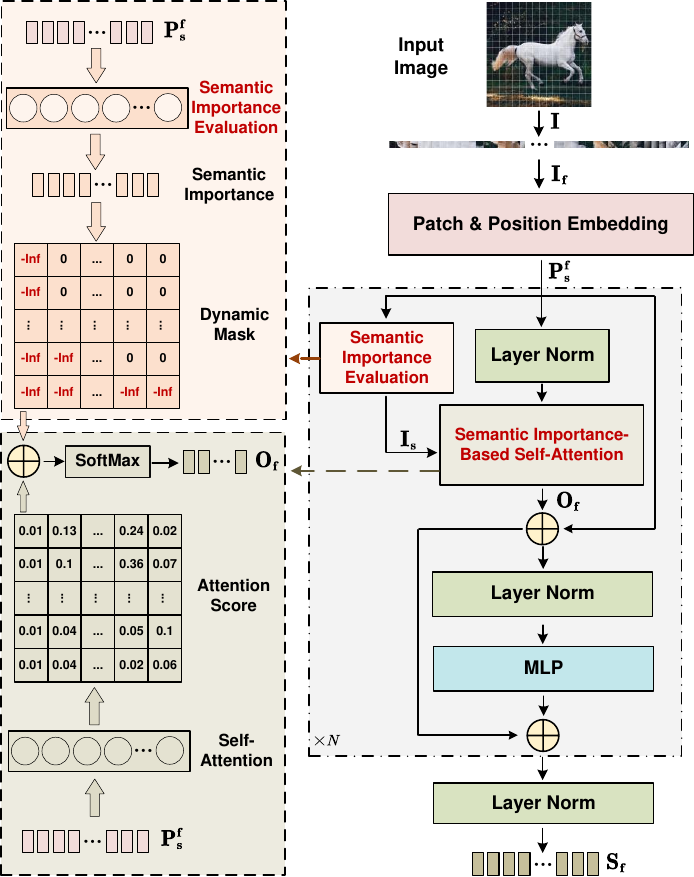}
        \caption{Architecture of the proposed fine-grained semantic extractor.}
        \label{fig: fine ViT}
\end{figure}

The architecture of the fine-grained semantic extractor is illustrated in Fig.~\ref{fig: fine ViT}. The image with semantic impairments, $\bm{{\rm I}}$, is segmented into small image patches, $\bm{{\rm I_f}}$. These patches are subsequently processed through both the patch and the position embedding layer to extract semantic information of each individual image patch, $\bm{{\rm P_s^f}}$, which may be represented as 
\begin{equation}
        \bm{{\rm P_s^f}} = \mathit{f}_{\bm{\psi }}(\bm{{\rm I_f}}),
\end{equation}
where $ \mathit{f}_{\bm{\psi}}(\cdot) $ is the patch and position embedding module with the trainable parameter set $\bm{\psi}$.

The semantic importance of image patches, $\bm{{\rm I_s}}$, is obtained with the semantic importance evaluation module, which can be described as 
\begin{equation}
        \bm{{\rm I_s}} = \mathit{f}_{\bm{\varepsilon  }}(\bm{{\rm P_s^f}}),
\end{equation}
where $ \mathit{f}_{\bm{\varepsilon}}(\cdot) $ is the semantic importance evaluation module with the trainable parameter set $\bm{\varepsilon}$.

Based on the semantic importance evaluation results, the element, $ m_{i,j} $, in position $(i, j)$ of the dynamic mask, $\bm{{\rm M_d}} $, is formulated as
\begin{equation}
        m_{i,j} = \left\{
                        \begin{array}{rcl}
                        0,  \quad & \{i,j\} \notin \mathbb{C},  \\
                        $-Inf$, \quad & \{i,j\} \in  \mathbb{C},
                        \end{array} \right
                        .
\end{equation}
where ${\rm -Inf}$ is the negative infinity, and $\mathbb{C}$ is the set of indexes corresponding to the top $k$ smallest value of semantic importance. 

The semantic importance-based self-attention score can be
\begin{equation}
        \bm{{\rm O_f}} = {\rm SoftMax}(\bm{{\rm A_t}} + \bm{{\rm M_d}})\bm{{\rm V}},
\end{equation}
where $\bm{{\rm A_t}}$ symbolizes the attention score, $\bm{{\rm V}}$ denotes the value, both of which are derived from the self-attention module, and $\bm{{\rm M_d}}$ is the dynamic mask originated through semantic importance evaluation. After conducting ${\rm SoftMax(\cdot)}$ operation, the attention scores of less significant regions are assigned a value of $0$, thereby directing the focus exclusively towards the most relevant and semantically meaningful areas.

The fine-grained semantic information is obtained by
\begin{equation}
        \bm{{\rm S_f}} = \mathit{f}_{\bm{\lambda}}(\bm{{\rm O_f}}),
\end{equation}
where $ \mathit{f}_{\bm{\lambda}}(\cdot) $ is the subsequent operations of fine-grained semantic encoder with the trainable parameter set $\bm{\lambda}$, which is shown in Fig.~\ref{fig: fine ViT}. 

\subsubsection{Coarse-Grained Semantic Extractor}

\begin{figure}[tbp]
        \centering
        \includegraphics[scale=0.5]{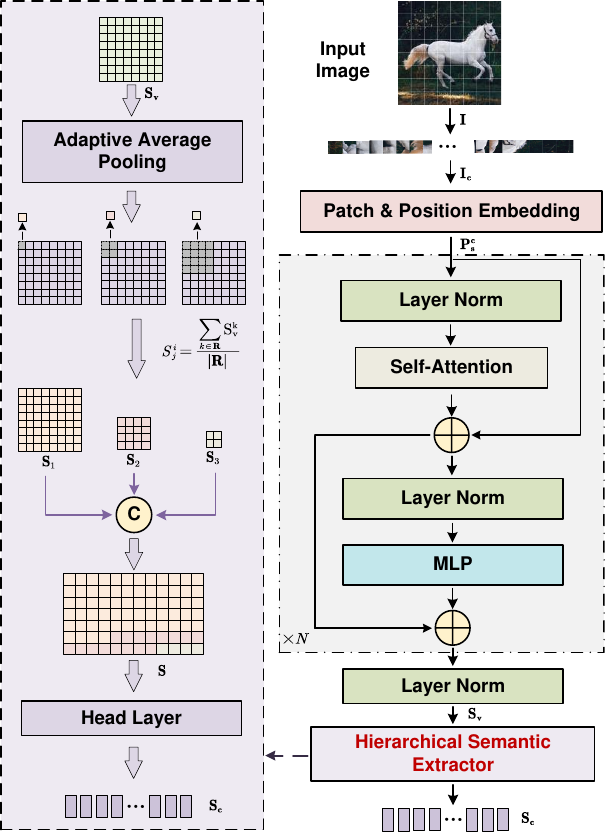}
        \caption{Architecture of the proposed coarse-grained semantic extractor.}
        \label{fig: coarse ViT}
\end{figure}

Fig.~\ref{fig: coarse ViT} depicts the proposed coarse-grained semantic extractor. Semantic information of image pathes is derived after passing the image patchs with large size, denoted as $\bm{{\rm I_c}}$, through the patch and the position embedding layer,  which may be expressed as 
\begin{equation}
        \bm{{\rm P_s^c}} = \mathit{f}_{\bm{\pi }}(\bm{{\rm I_c}}),
\end{equation}
where $ \mathit{f}_{\bm{\pi }}(\cdot) $ is the patch and position embedding module with the trainable parameter set $\bm{\pi}$.

The current semantic information is calculated as
\begin{equation}
        \bm{{\rm S_v}} = \mathit{f}_{\bm{o}}(\bm{{\rm P_s^c}}),
\end{equation}
where $ \mathit{f}_{\bm{o}}(\cdot) $ represents the remaining operations of ViT with the trainable parameter set $\bm{o}$.

Ultimately, a novel hierarchical semantic extractor is introduced to enhance the robustness of semantic information by incorporating pooling techniques~\cite{adaptive_pooling}. The $j_{th}$ element of level-$i$ semantic information, $\bm{{\rm S^i}}$, can be expressed as 
\begin{equation}
        S^i_j = \frac{\sum\limits_{k \in \bm{{\rm R}}} {\rm S_v^k}} {|\bm{{\rm R}}|},
\end{equation}
where $\bm{{\rm R}}$ is the set of level-$i$ pooling regions and ${\rm S_v^k}$ is $k_{th}$ value of the $\bm{{\rm S_v}}$ in $\bm{{\rm R}}$. We implement a $3$-level pooling for feature refinement. 

The semantic information, $\bm{{\rm S}}$, is obtained by conducting concatenation, which is represented as
\begin{equation}
        \bm{{\rm S}} = {\rm concat}( \bm{{\rm S^1}},  \bm{{\rm S^2}},  \bm{{\rm S^3}}),
\end{equation}
where $\bm{{\rm S^1}}$, $\bm{{\rm S^2}}$, $\bm{{\rm S^3}}$ is the semantic information extracted from the pooling process, while ${\rm concat(\cdot)}$ is the concatenate operation, which is marked as \textcircled{c} in Fig.~\ref{fig: coarse ViT}.

The semantic information, $\bm{{\rm S_c}}$, extracted by the coarse-grained semantic extractor is represented as
\begin{equation}
        \bm{{\rm S_c}} = \mathit{f}_{\bm{\nu }}(\bm{{\rm S}}),
\end{equation}
where $\mathit{f}_{\bm{\nu }}(\cdot)$ is the head layer which consists of a linear layer and a Sigmoid function, and the trainable parameter set $\bm{\nu }$.

\subsubsection{Cross-Attention-Based Semantic Fusion Module}
After developing both the coarse-grained and the fine-grained semantic extractors, the next step is to blend the semantics derived from these dual branches. 

The coarse-grained semantics, $\bm{{\rm S_c}}$, and the fine-grained semantics, $\bm{{\rm S_f}}$, are further processed by two independent projection layer for subsequent operations. The processed coarse-grained semantics, $\bm{{\rm S_c'}}$, can be represented as 
\begin{equation}
        \bm{{\rm S_c'}} = \mathit{f}_{\bm{\vartheta }}(\bm{{\rm S_c}}),
\end{equation}
where $\mathit{f}_{\bm{\vartheta }}(\cdot)$ is the coarse-grained projection layer with the trainable parameter set $\bm{\vartheta }$.

The processed fine-grained semantics, $\bm{{\rm S_f'}}$, can be 
\begin{equation}
        \bm{{\rm S_f'}} = \mathit{f}_{\bm{\theta  }}(\bm{{\rm S_f}}),
\end{equation}
where $\mathit{f}_{\bm{\theta  }}(\cdot)$ is the fine-grained projection layer with the trainable parameter set $\bm{\theta  }$.

Based on the processed coarse-grained and fine-grained semantic information, we design a module with the cross-attention mechanism for feature fusion. The query, $\bm{{\rm Q}}$, the key, $\bm{{\rm K}}$, and the value, $\bm{{\rm V}}$, can be formulated as
\begin{equation}
        \bm{{\rm Q}} = \mathit{f}_{\bm{\kappa}}(\bm{{\rm S_f'}}),
\end{equation}
\begin{equation}
        \bm{{\rm K}} = \mathit{f}_{\bm{\xi }}(\bm{{\rm S_f'}}),
\end{equation}
\begin{equation}
        \bm{{\rm V}} = \mathit{f}_{\bm{\varpi  }}({\rm concat}(\bm{{\rm S_f'}}, \bm{{\rm S_c'}})),
\end{equation}
where $\mathit{f}_{\bm{\kappa}}(\cdot)$, $\mathit{f}_{\bm{\xi }}(\cdot), \mathit{f}_{\bm{\varpi }}(\cdot)$ represents the Q projection layer with the trainable parameter set $\bm{\kappa }$, the K projection layer with the trainable parameter set $\bm{\xi }$, and the V projection layer with the trainable parameter set $\bm{\varpi }$.

The multi-grained semantics is obtained by 
\begin{equation}
        \bm{{\rm S_{m}}} = \mathit{f}_{\bm{\rho }}
        (\mathit{{\rm softmax}}(\bm{{\rm Q}} \cdot \bm{{\rm K}} ^ T) \cdot \bm{{\rm V}}),
\end{equation}
where $\mathit{f}_{\bm{\rho  }}(\cdot)$ is the projection layer with the trainable parameter set $\bm{\rho  }$ to generate the final semantics. 


\subsection{Channel Codec and Semantic Decoder}

After extracting the multi-scale semantic information, the transmitter applies channel encoding to produce the transmitted signal, $\bm{{\rm T_x}}$, enhancing its robustness against physical impairments, which is given by
\begin{equation}
        \bm{{\rm T_x}} = \mathit{f}_{\bm{\epsilon }}(\bm{{\rm S_{m} }}),
\end{equation}
where $\mathit{f}_{\bm{\epsilon}}(\cdot)$ represents the channel encoder which consists of linear layers and the trainable parameter set $\bm{\epsilon}$.

The received signal, $\bm{{\rm R_x}}$, undergoes processing by the channel decoder to recover the multi-scale semantic information, which is expressed as
\begin{equation}
        \bm{{\rm \hat{S}_{m}}} = \mathit{f}_{\bm{\gamma}}(\bm{{\rm R_x }}),
\end{equation}
where $\mathit{f}_{\bm{\gamma}}(\cdot)$ represents the channel decoder which consists of linear layers and the trainable parameter set $\bm{\gamma}$.

Ultimately, semantic decoder, which consists of ResBlocks, Attention layers, and a projection layer, generates the purified image to eliminate semantic impairments. The output of the semantic decoder is represented as follows:
\begin{equation}
        \bm{{\rm \hat{I}}} = \mathit{f}_{\bm{\varrho  }}(\bm{{\rm \hat{S}_{m}}}),
\end{equation}
where $\mathit{f}_{\bm{\varrho  }}(\cdot)$ represents the semantic decoder with the trainable parameter set $\bm{\varrho}$.

\subsection{Loss Function}
We introduce the loss function to train the robust semantic communication system, which is defined as
\begin{equation}
        \mathcal{L}_{total} = \mathcal{L}_{CE}(\bm{{\rm I_u}}, \bm{\hat{{\rm I}}}) + \alpha \cdot \mathcal{L}_{MSE}(\bm{{\rm T_x}}, \bm{{\rm R_x}}), 
\end{equation}
where $\bm{{\rm I_u}}$ is the uncorrupted image, $\alpha$ is the predefined weight parameter of a positive value used to adjust the weights of the two components in the loss function, $\mathcal{L}_{CE}(\cdot)$ represents the cross-entropy loss, and $\mathcal{L}_{MSE}(\cdot)$ represents the mean squared error loss. The first part of the loss function aims to mitigate semantic impairments, while the second part is designed to address physical channel impairments. 


\section{Numerical Results}

\subsection{Semantic Impairment Dataset}

We adopt CIFAR10~\cite{cifar10} in our experiments. CIFAR10 is a image classification dataset comprising $50, 000$ images for training and $10, 000$ images for testing. We choose PGD~\cite{PGD} for creating adversarial semantic impairments based on the downstream task models~\cite{carmon2019unlabeled} to construct the semantic impairment dataset for image classification.

\subsection{Baseline Models and Simulation Settings}
The proposed method is compared with a series of existing methods. The first one is the semantic communication system based on the Vision Transformer approach~\cite{vit}. The second one is the semantic communication system that utilizes the UNet backbone architecture~\cite{unet} as semantic codec. The third one is the VQ-DeepSC~\cite{fu2023vector}, which is a knowledge base assisted semantic communication system for image transmission. Moreover, a traditional communication system is considered, which employs the BPG as source codec, the LDPC for channel codec, and the $16$ QAM for modulation. 

\subsection{Performance Metrics}

\subsubsection{PSNR}
PSNR is the metric used for evaluating image quality, which is given by
\begin{equation}
        {\rm PSNR}(\bm{{\rm I_g}}, \bm{{\rm I_r}}) =  10 {\rm log}_{10}(\frac{V_{max}}{{\rm MSE}(\bm{{\rm I_g}}, \bm{{\rm I_r}})}),
\end{equation}
where $\bm{{\rm I_g}}$, $\bm{{\rm I_r}}$ is the groud-truth image and the received image for evaluation respectively, $V_{max}$ is the maximum pixel value, and ${\rm MSE}(\cdot)$ is the function of the mean squared error. 


\subsubsection{LPIPS}
Distinct from the PSNR, the LPIPS evaluate similarity in a semantic space. The LPIPS score is obtained by averaging the similarity scores across patches, which is represented as
\begin{equation}
        {\rm LPIPS}(\bm{{\rm I_g}}, \bm{{\rm I_r}}) = \sum\limits_{l}  \frac{1}{H_l W_l}  \sum\limits_{i, j} (\bm{{\rm w_l}} \bigodot {\Vert \bm{{\rm F_g^{l}}} - \bm{{\rm F_r^{l}}} \Vert}_{2}^2),   
\end{equation}
where $H_l$, $W_l$ is the height and weight of the feature map in $l_{th}$ patch,  $\bm{{\rm w_l}}$ is the weights for different patches, $\bigodot$ represents the element-wise product, $\bm{{\rm F_g^{l}}}$ and $\bm{{\rm F_r^{l}}}$ is the $l$-level feature map of the groud-truth image and the received image respectively. 

\subsubsection{Accuracy}
Accuracy is a widely employed performance metric in classifications to quantify the ability to correctly classify samples, which is given by
\begin{equation}
        {\rm ACC} = \frac{{\rm TP + TN}}{{\rm TP + TN + FP + FN}}, 
\end{equation}
where ${\rm TP}$, ${\rm TN}$, ${\rm FP}$, and ${\rm FN}$ represent the number of true positive, true negative, false positive, and the false negative samples, respectively, which correspond to the outputs of classification by the pretrained model~\cite{carmon2019unlabeled}.

\subsection{System Performance}

In this section, we present our experimental results. To demonstrate the effectiveness of the proposed method, we conducted experiments under different signal-to-noise ratios (SNRs) and various levels of ISIIs.

\subsubsection{System Performance Versus SNR}

\begin{figure*}[tbp]
	\centering
	\subfigure{
	\includegraphics[width=0.6\linewidth]{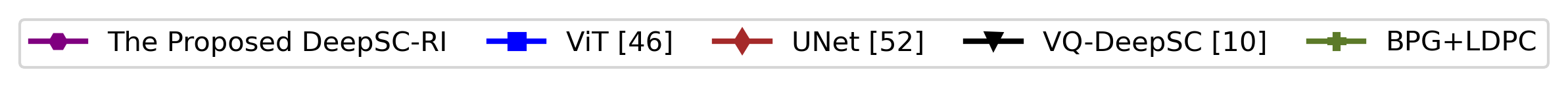}
	}
	\vspace{-0.5cm}

	\centering
	\setcounter{subfigure}{0}
	\subfigure[LPIPS Versus SNR]{
		\includegraphics[width=0.2\linewidth]{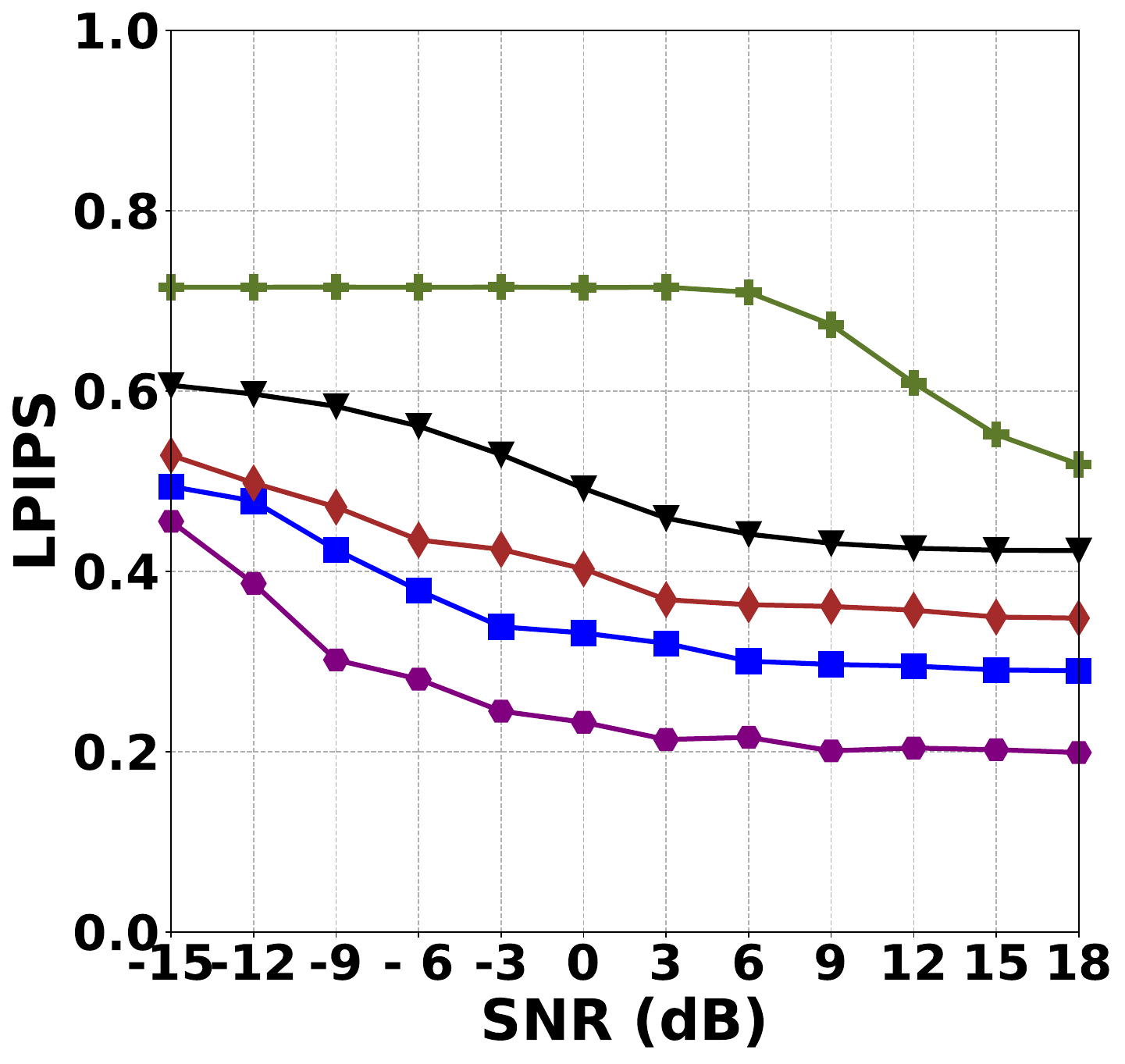}
		\label{fig: rician_0.22_lpips}
		}
	\quad
	\subfigure[PSNR Versus SNR]{
		\includegraphics[width=0.2\linewidth]{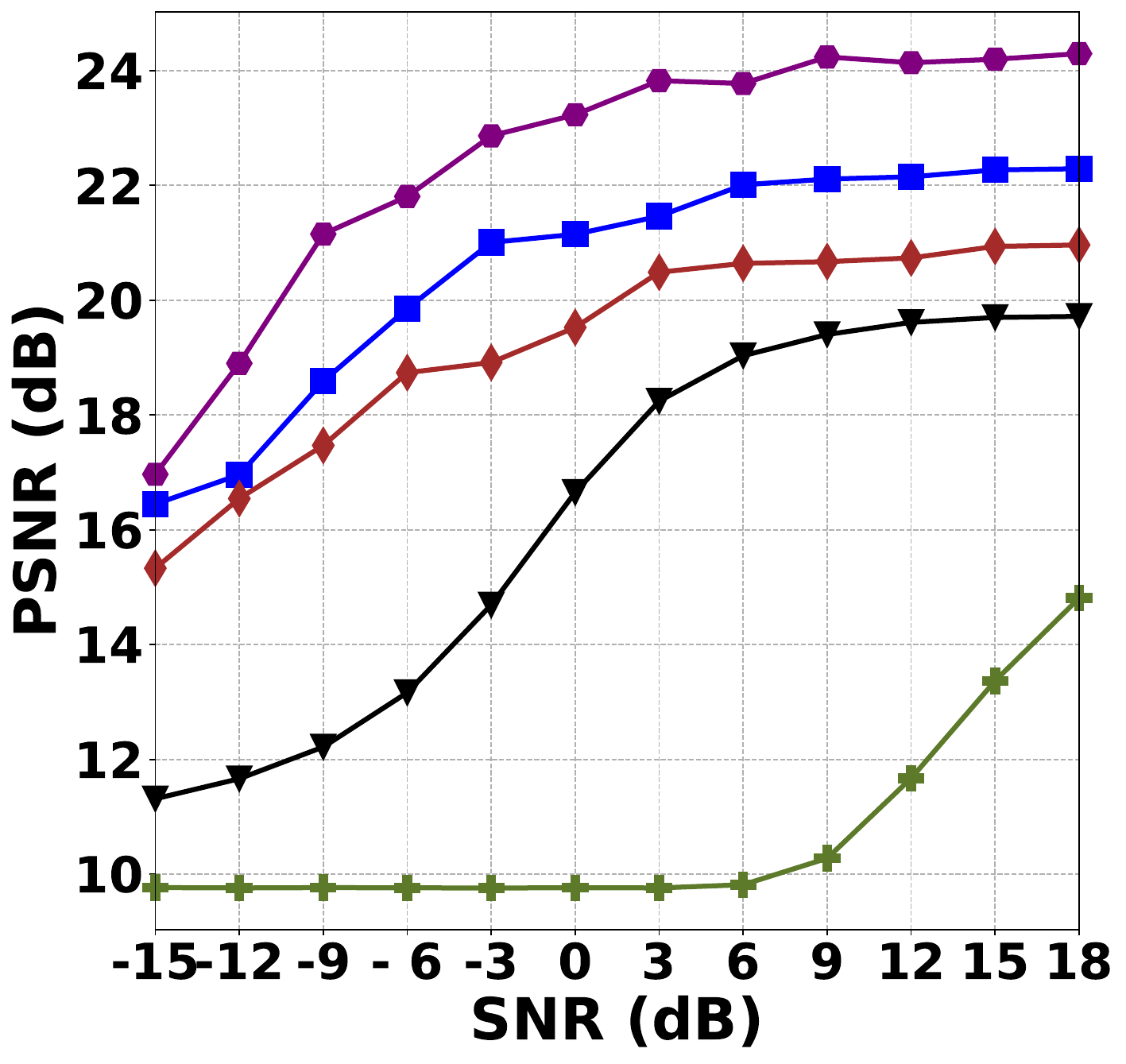}
		\label{fig: rician_0.22_psnr}
		}
	\quad
	\subfigure[ACC Versus SNR]{
		\includegraphics[width=0.2\linewidth]{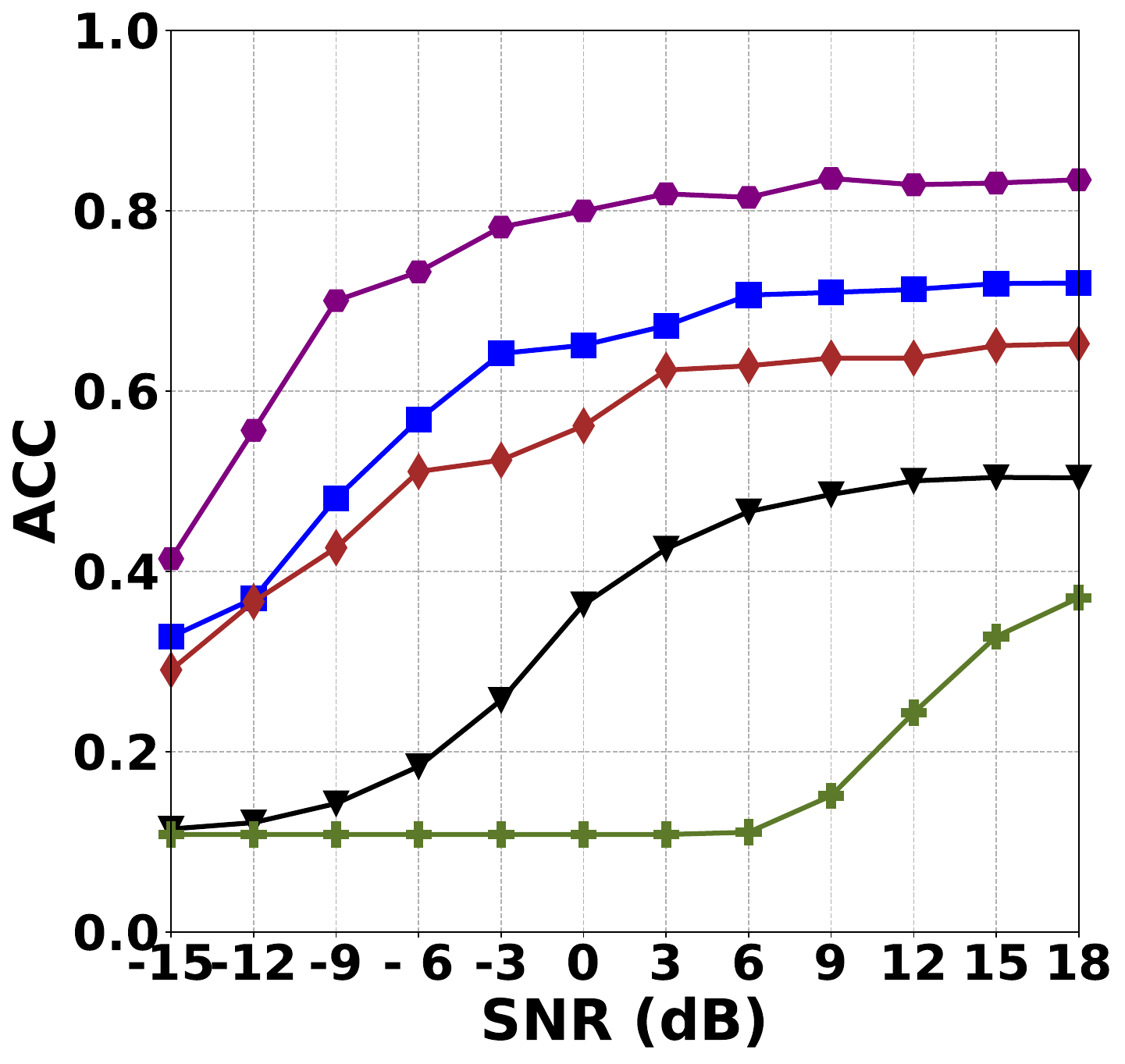}
		\label{fig: rician_0.22_acc}
		}
	\quad
	\caption{System performance under Rician fading channels versus SNR with CIFAR10. }
	\label{fig: Rician_1}
\end{figure*}

We conducted experiments under Rician fading channels, the results are illustrated in Fig.~\ref{fig: Rician_1}. A noticeable observation is that the traditional communication system suffers a significant performance decline. At an SNR of $18$ dB, the classification accuracy of the traditional method falls below $40\%$, highlighting its vulnerabilities to semantic impairments. In contrast, semantic communication systems demonstrate a divergent performance. Despite some degradations at low SNRs, the semantic communication systems consistently maintain excellent semantic fidelity. This resilience is mainly due to their inherent capability for semantic understanding and interpretation, which enables them to counteract disturbances caused by physical channels and semantic impairments.

Besides, as witnessed by the three metrics, the proposed DeepSC-RI demonstrates superior performance under Rician fading channels, which indicates that the proposed multi-scale semantic extractor is eminently suitable for eliminating semantic impairment and remaining robust under complex transmission environment. 

\subsubsection{System Performance Versus ISII}

To explore the correlation between semantic fidelity and \textit{image semantic impairment intensity}, we conducted experiments at varying ISIIs. The adopted ISIIs vary from $0.2$ to $0.8$ with an interval of $0.1$, while maintaining a constant SNR of $18$ dB.

\begin{figure*}[tbp]
	\centering
	\subfigure{
	\includegraphics[width=0.6\linewidth]{legend.png}
	}
	\vspace{-0.5cm}

	\centering
	\setcounter{subfigure}{0}
	\subfigure[LPIPS Versus ISII]{
		\includegraphics[width=0.2\linewidth]{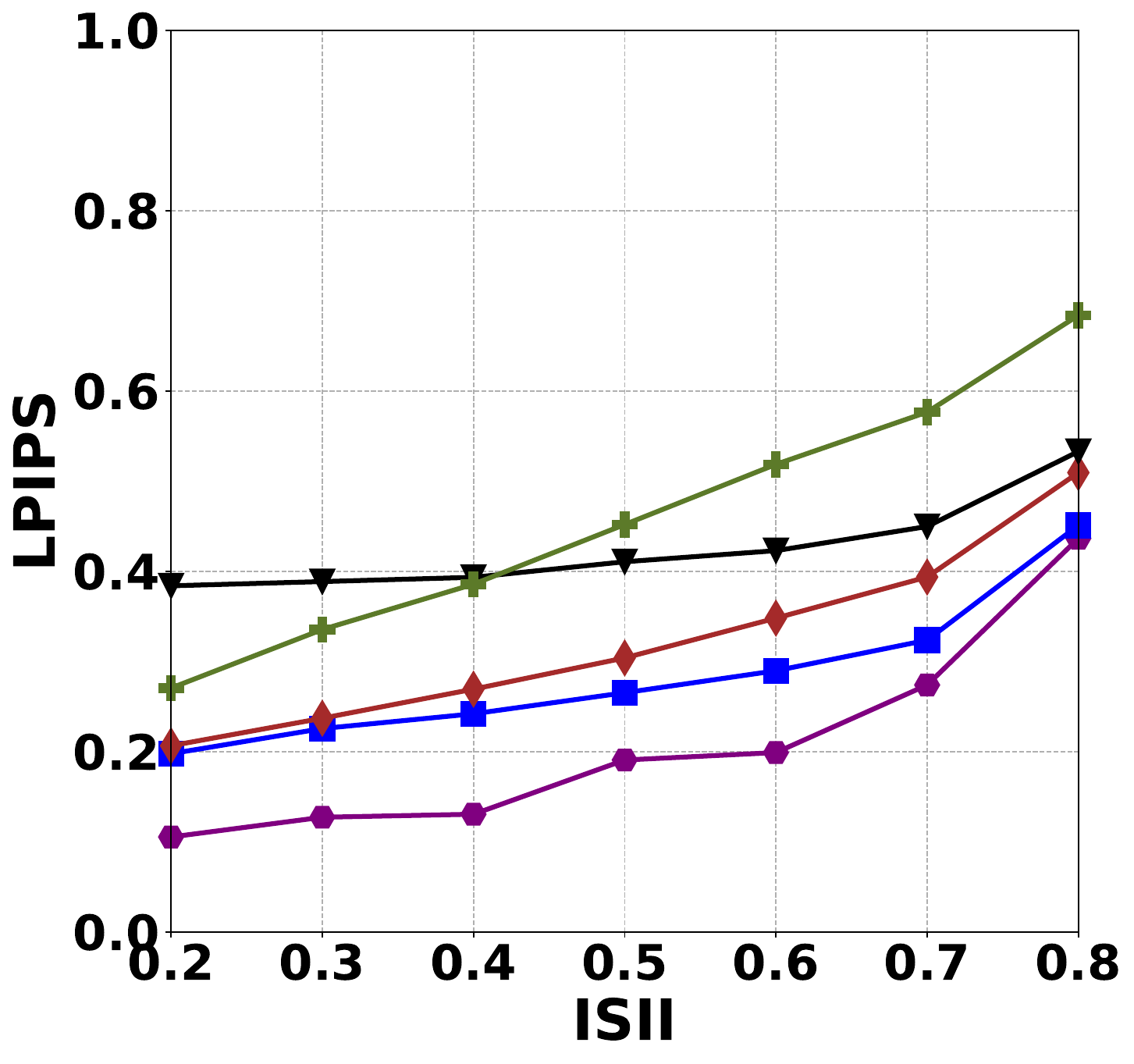}
		\label{fig: rician_lpips}
		}
	\quad
	\subfigure[PSNR Versus ISII]{
		\includegraphics[width=0.2\linewidth]{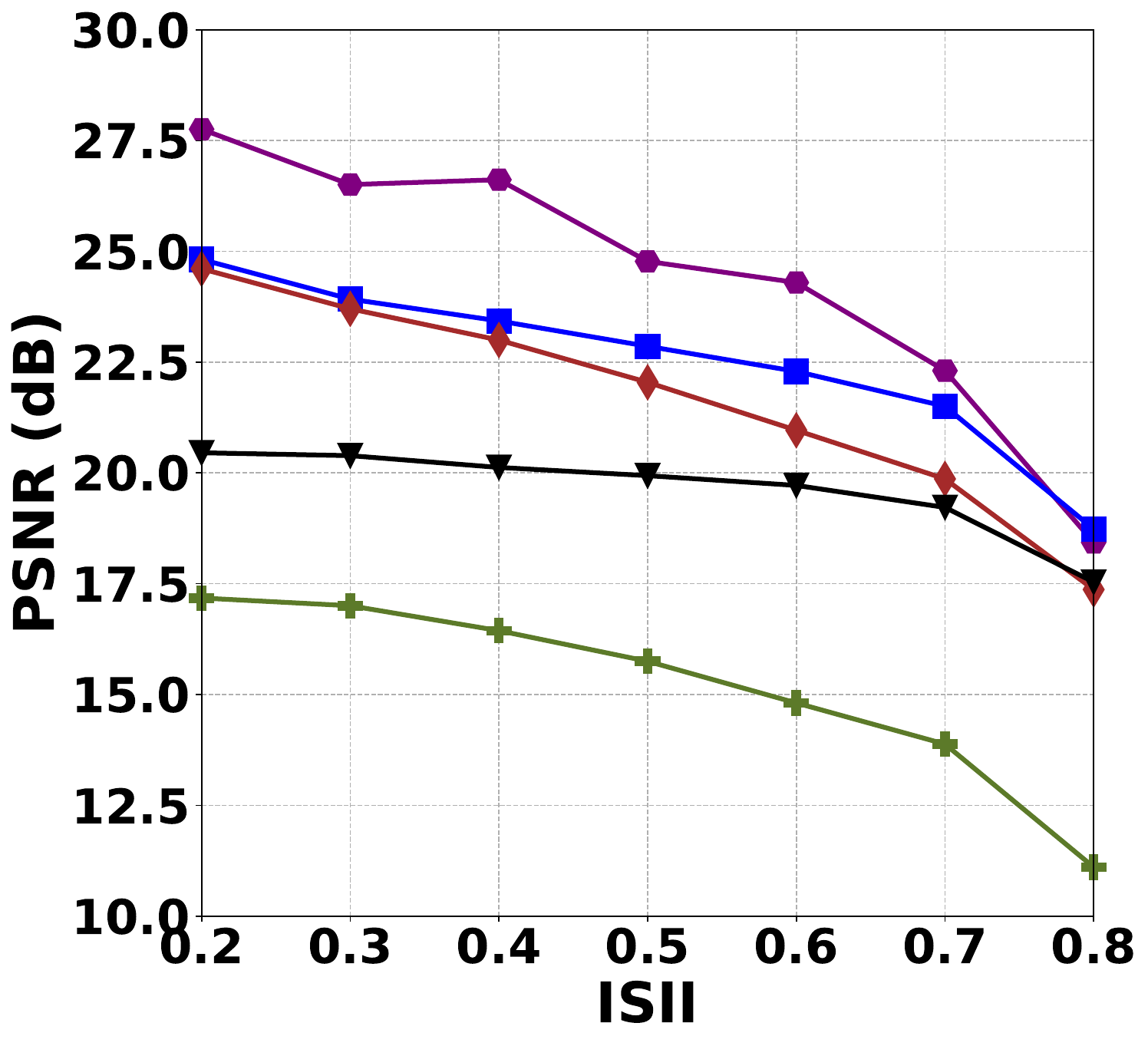}
		\label{fig: rician_psnr}
		}
	\quad
	\subfigure[ACC Versus ISII]{
		\includegraphics[width=0.2\linewidth]{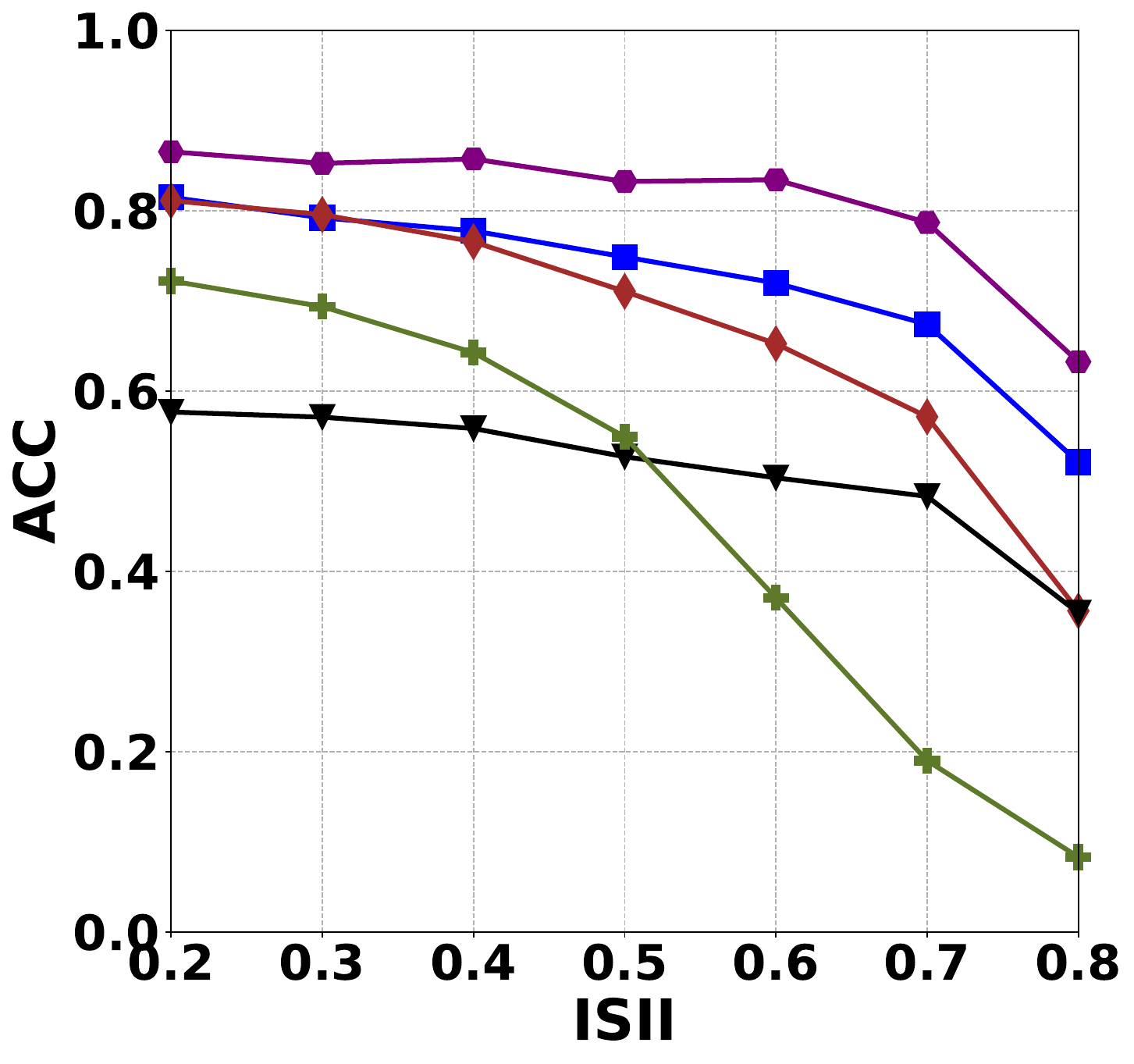}
		\label{fig: rician_acc}
		}
	\quad
	\caption{System performance under Rician fading channels versus ISII with CIFAR10. }
	\label{fig: Rician}
\end{figure*}

The results under Rician fading channels are presented in Fig.~\ref{fig: Rician}. The traditional method employing BPG and LDPC method suffers performance degradation across various ISIIs, while the semantic communication systems experience minor reductions but yield remarkable robustness in classification accuracy. As illustrated in Fig.~\ref{fig: rician_acc}, at an ISII of $0.3$, the semantic communication systems employing the ViT and the UNet achieve classification accuracy of around $80\%$. Conversely, the traditional method faces a substantial decline in classification accuracy, falling to nearly $70\%$.

Moreover, it is evident that as ISII intensifies, the semantic fidelity for all systems exhibits obvious decreases. This observation indicates the negative impact of semantic impairments on semantic communications, emphasizing the vital necessity of developing robust systems to against semantic impairments.

Additionally, the semantic communication systems demonstrate enhanced robustness to semantic impairments in contrast to the traditional
communication system. This observation validates the adaptability of semantic communications under complex transmission scenario, especially in environments characterized by high levels of semantic impairment. 

Furthermore, it is remarkable that although the semantic fidelity of all systems diminishes as semantic impairment escalates, our proposed DeepSC-RI distinctly achieves superior performance, especially in classification accuracy, which further validate the robustness of the proposed system.



\section{Conclusion}

This paper addresses the issue of semantic impairments in image, which is defined as adversarial perturbations at the source. Firstly, we have introduced a novel metric to quantify the intensity of semantic impairment. Afterwards, we have constructed a semantic impairment dataset for validation. Furthermore, we have proposed a semantic communication system, namely DeepSC-RI, to eliminate semantic impairment, which is built upon a multi-scale ViT. The multi-scale ViT consists of two distinct branches: coarse-grained and fine-grained semantic extractor. The fine-grained branch incorporates a semantic importance evaluation module that identifies crucial semantics, while the coarse-grained branch employs a hierarchical strategy to progressively generate coarse-grained semantics. The semantic fusion module combines insights from both branches using a cross-attention mechanism. The experimental results demonstrate that the propsoed architecture can improve the semantic fidelity of the system by eliminating semantic impairments. 

\section*{Acknowledgment}
This work was supported by the National Natural Science Foundation of China (NSFC 61925105, 62293484, 62227801) and Shanghai Municipal Science and Technology Major Project (Grant No.2018SHZDZX04).

\bibliographystyle{IEEEtran}
\bibliography{reference.bib}

\end{document}